\documentclass[aps,prl,twocolumn,showpacs]{revtex4} 
\usepackage{graphicx}
\usepackage{dcolumn}
\usepackage{bm}
\usepackage{epstopdf}
\usepackage{amstext} 
\usepackage{fixmath}

\begin{document}

\newcommand{\EW}{E{\"o}t-Wash}
\newcommand{\w}{$\omega$}
\newcommand{\m}{$\mu$}

\title{New Test of the Gravitational $1/r^2$ Law at Separations down to 52 $\mu$m}

\author{J.\,G.~Lee}
\author{E.\,G.~Adelberger}
\email{eadelberger@gmail.com}
\author{T.\,S.~Cook}
\altaffiliation{Present Address:  Micro Encoder Inc., Kirkland, WA 98034 USA}
\author{S.\,M.~Fleischer}
\altaffiliation{Present Address: Department of Mechanical Engineering, University of Washington, Seattle, WA 98195, USA.}
\author{B.\,R~Heckel.}
\affiliation{Center for Experimental Nuclear Physics and Astrophysics, Box 354290,
University of Washington, Seattle, Washington 98195-4290 USA} 
\begin{abstract}
We tested the gravitational $1/r^2$
 law using a stationary torsion-balance detector and a rotating attractor containing test bodies with both 18-fold and 120-fold azimuthal symmetries that simultaneously tests the $1/r^2$ law at two different length scales. 
We took data at detector-attractor separations between $52~\mu$m and 3.0 mm. Newtonian gravity gave an excellent fit to our data,
 limiting with 95\% confidence any gravitational-strength Yukawa interactions to ranges $< 38.6~\mu$m.
\end{abstract}
\maketitle
Testing gravity at the shortest attainable distances is interesting for many reasons. 
String theory's unifcation of gravity with the other 3 fundamental interactions inherently involves extra gravitational space dimensions as well as many nominally-massless scalar particles (dilaton and moduli). Both of these features would violate the gravitational inverse-square law (ISL) \cite{ar:98,ad:03}, as would a second, heavy graviton\cite{ao:16}.  New phenomena could occur below the length scale associated with dark energy $\lambda_{\rm d}\!=\!\sqrt[4]{\hbar c/\rho_{\rm d}}\!\approx\! 85~\mu$m where $\rho_{\rm d}\!\approx \!3.8$ keV/cm$^3$ is the observed density of dark energy\cite{su:04,ka:07}. Suggestions that dark matter may consist of ultra-low-mass scalar and vector bosons\cite{ne:11,gr:16}, whose exchange interaction would violate the ISL, provide
further motivation for exploring this regime. 
It is customary to interpret ISL data as constraining an additional Yukawa interaction
\begin{displaymath}
V(r)=V_N(r) [1+\alpha \exp({-r/\lambda})]~,
\end{displaymath}
where $V_N(r)$ is the familiar Newtonian potential. This form is obviously valid for scalar or vector boson-exchange interactions and is a reasonable approximation for the effects of extra dimensions as long as the minimum separation attained in the experiment is greater than the size of the largest extra dimension\cite{extra}.

Precise studies of gravity at length scales below $100~\mu$m are challenging because the tiny gravitational forces exerted by appropriately-sized test bodies are easily ``polluted'' by extraneous effects.  
Here we report results from two latest generations of the E{\"o}t-Wash rotating-attractor torsion-balance ISL tests. In these tests,  harmonic torques,  exerted on a detector pendulum by a rotating attractor, are studied as functions of separation $s$ between the facing surfaces of the detector and attractor test bodies. Our new device offers significant improvements over those used previously\cite{ho:01,ho:04,ka:07,ad:07}. The new test-body design (see Fig.~1) has both 18-fold and 120-fold azimuthal symmetries and tests the ISL at two different length scales at once. The 50\%-transparent hole pattern maximizes the signals for a given test-body diameter. Furthermore, the Fourier-Bessel expansion provides nearly-analytic (a single numerical integration) solutions for Newtonian, Yukawa and dipole-dipole torques\cite{co:13,te:15}. In our device, the primary science signals are torques varying at $18\omega$ and $120\omega$ where $\omega$ is the attractor rotation frequency. Figure~1 shows the predicted Newton and Yukawa torques as functions of $s$. 
%
\begin{figure}[!b]
	 \includegraphics[width=0.47\textwidth]{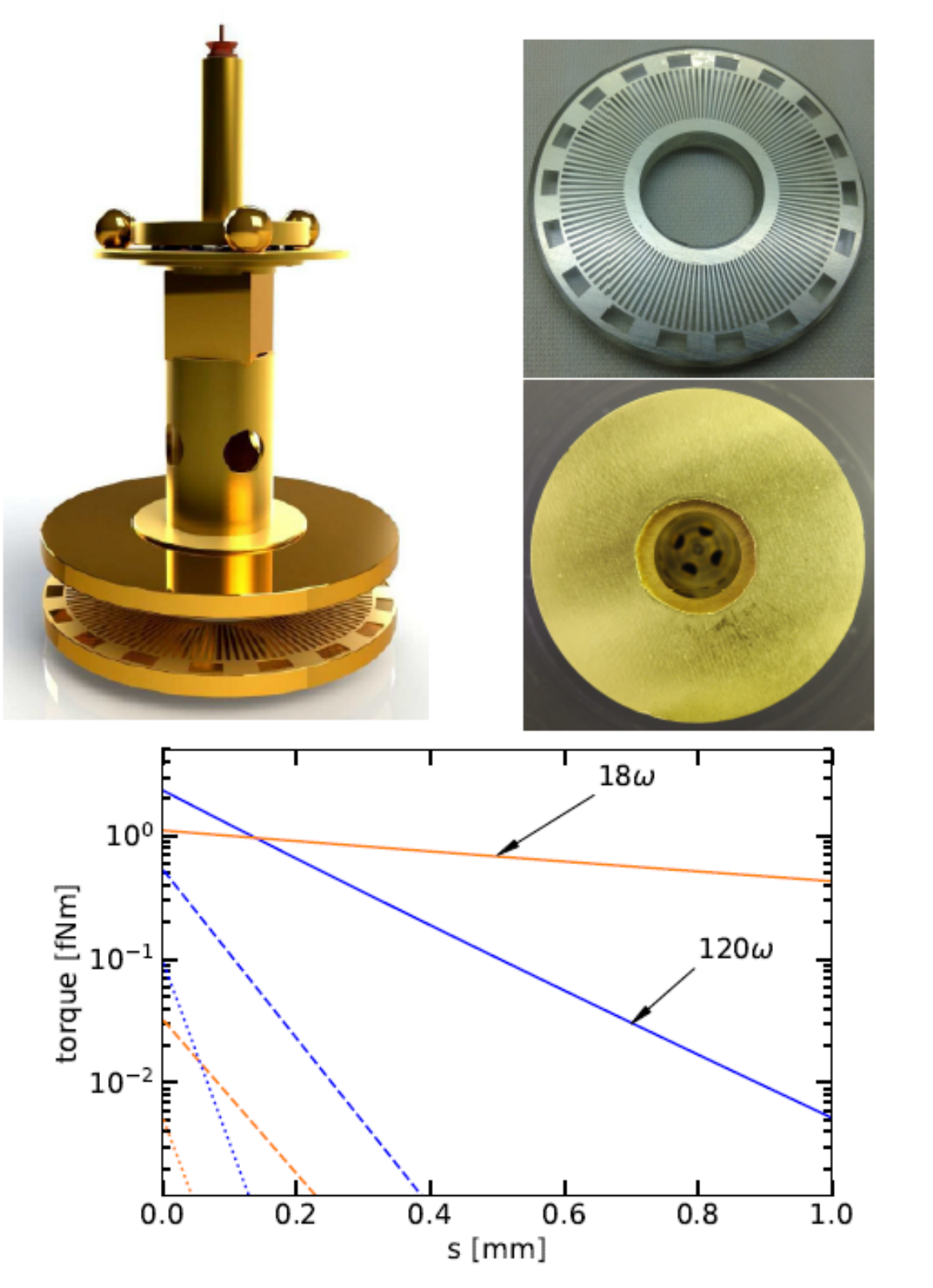}
\caption{(color online)  {\bf Top left:} detector and rotating attractor rendered with their separation much larger than actually used. An electrostatic shield that surrounds the detector and isolates it from the attractor is not shown. {\bf Top right:} photos of glued generation 2 test body before and after gold coating. The hole pattern diameter is 52 mm. {\bf Bottom plot:} predicted Newtonian torques (solid lines). The dashed and dotted lines are $\alpha\!=\!1$ Yukawa torques with $\lambda=70$ and 30 $\mu$m, respectively}
\label{fig1}	
\end{figure} 
Note that as $s$ increases, the torques decay exponentially with scale lengths inversely proportional to the azimuthal symmetry number $n$. 
We calibrated the torque scale (see Fig.~\ref{fig2}) by the gravitational interaction between the 3 small spheres on the detector and 3 large spheres on an external turntable. The separation between the 2 sets of spheres was
comparable to those used in measuring Newton's constant\cite{ro:17} and in a regime where independent experiments\cite{ho:85} have verified the $1/r^2$ law at the $10^{-3}$ level; our work can be viewed as percent-level measurements of $G_N$ at separations down to about $50\,\mu$m.

Our raw data consist of torque measurements at a set of 3-dimensional displacements $\vec{\zeta}=(x,y,s)$ between the detector and attractor test bodies. Each data point comprised $\theta$ (an autocollimator measurement of the detector twist angle), $\phi$ (the turntable angle from a high-resolution encoder), the capacitance between the detector and the electrostatic shield (a key element in determining $s$), plus a dozen other parameters such as apparatus tilts, various temperatures, {\em etc}. Because the Fourier-Bessel hole pattern repeats every 60 degrees the data streams were cut into 60 degree segments typically containing 680 points.  The 
$\theta(\phi)$ data in each cut were fit with harmonic terms and low-order polynomial drift\cite{ho:04,ka:07}. The data-taking cadence and $\omega$ were set so that each cut contained integral numbers of data points and free-torsional oscillations and that the $18\omega$ and $120\omega$ signals lay in low-noise regions of the torque power spectrum (see bottom panel in Fig.~\ref{fig2}). 
Harmonic torques $N_{n\omega}$ were inferred using $N_{n\omega}=\tilde{\theta}_{n\omega}I\omega_0^2$ where $\tilde{\theta}_{n\omega}$
was the harmonic amplitude corrected for pendulum inertia plus electronic and digital-filter\cite{co:13} time constants, $I$ ($91.7\,$g$\,$cm$^2$ in  generation 2) was the detector's rotational inertia computed from a detailed numerical model, and 
$\omega_0 \approx .0184\,{\rm s}^{-1}$ was the detector's free-oscillation frequency. (Uncertainties in $I$ have no effect on our results because $I$ appears in both the Fourier-Bessel and calibration-sphere torques.) Electrostatic ``patch'' effects\cite{patch} altered $\omega_0$ at small $s$ so, before and after each science run, $\omega_0$ was measured  in ``sweep runs'' where the free-oscillation amplitude (typically $\sim 4\,\mu$rad) was increased to about $20\,\mu$rad; this gave more precise values for $\omega_0$ and also provided corrections for small nonlinearities in the autocollimator angle scale. 
As shown in Fig.~\ref{fig3}, accurate values for the displacement, $\vec{\zeta}$, were obtained with the aid of micrometers on the x,y,z stage that supported the torsion fiber. Measurements of the $120\omega$ gravitational torques as functions of x and y gave the horizontal displacement, while electrical capacitances as functions of z were used to obtain the vertical displacement .  
%
\begin{figure}[t]
\includegraphics[width=.45\textwidth]{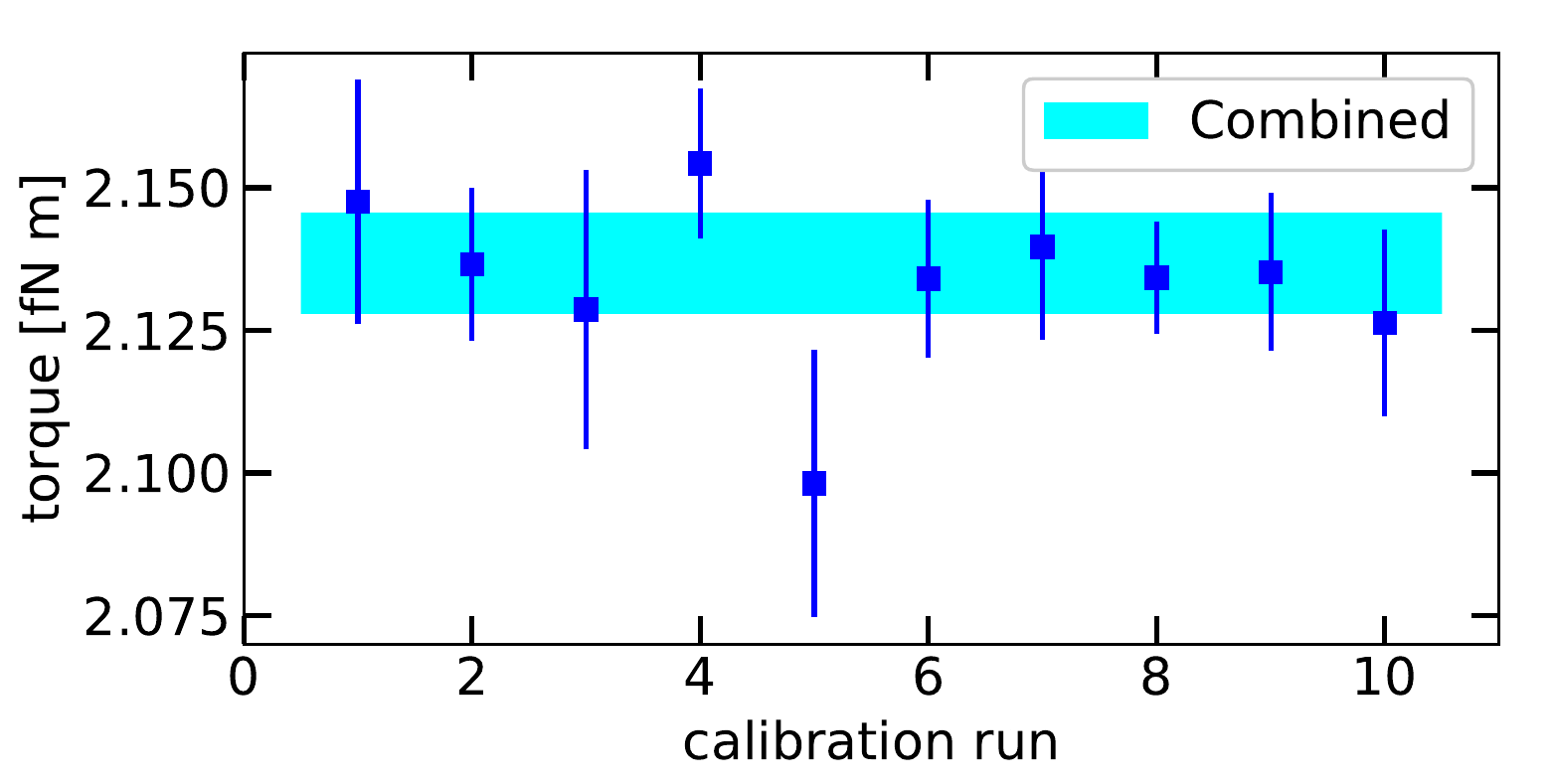}      
\includegraphics[width=.45\textwidth]{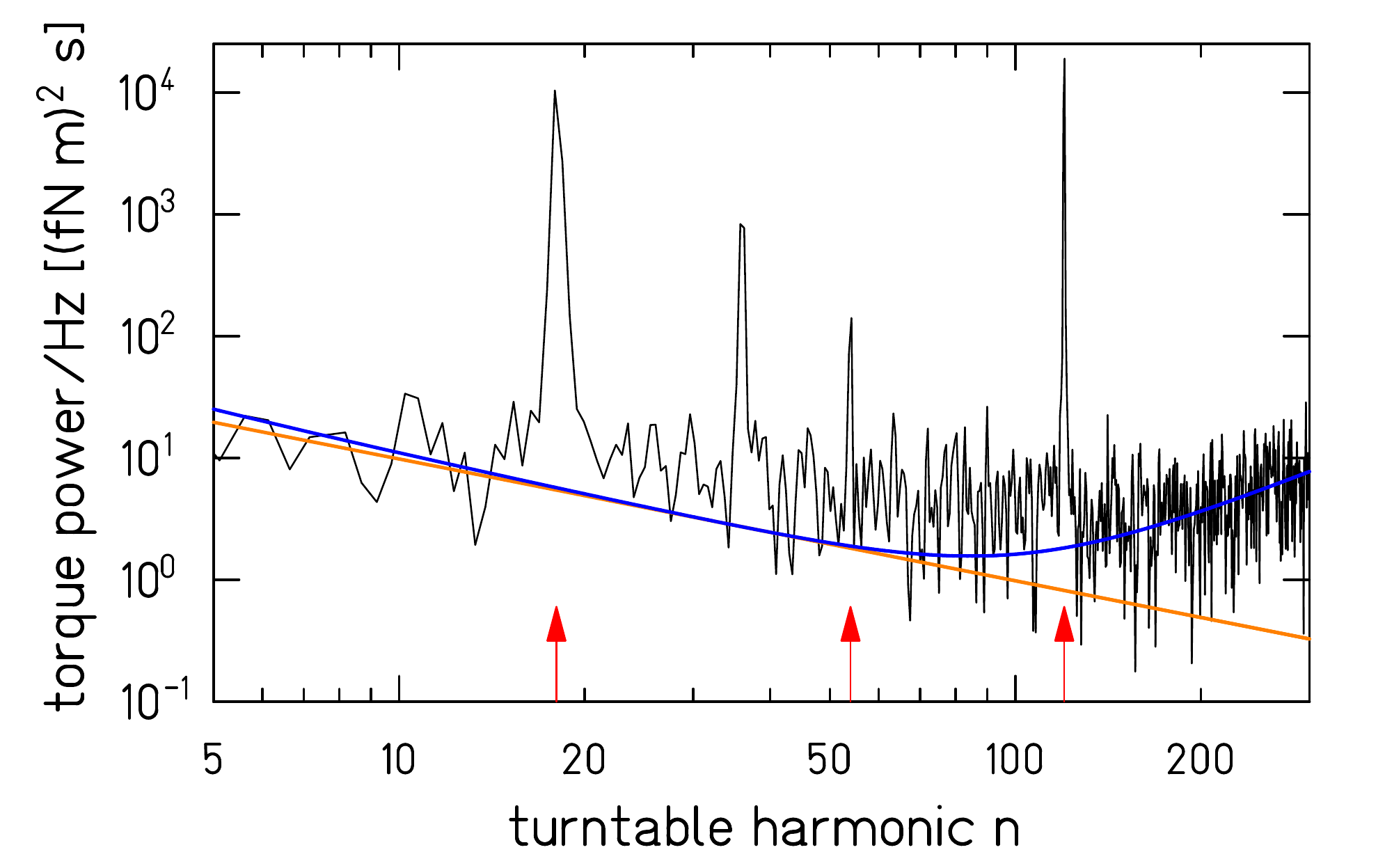}      
\caption{(color online)  Generation 2 data. {\bf Top plot:} absolute calibration of the torque scale. Three 1.137 kg spheres on an external calibration turntable rotating at $\omega_c$ applied a $3\omega_c$ gravitational torque on three 0.4816 g spheres on the detector. The detector and calibration spheres were equally spaced on 16.48 mm and 19.05 cm radius circles, respectively. 
 ($\omega_c$ was chosen to put the $3\omega_c$ signal's frequency close to those of the $120\omega$ signals in our science runs.) The result, $(2.137\pm 0.009)\,$fN\,m,  is based on the nominal autocollimator calibration. The expected gravitational torque is $(2.112\pm0.005)$~fN\,m where the error arises from uncertainties in the positions and masses of the spheres.  The expected/nominal   ratio $(\gamma\!=\!0.988\pm0.005)$ provided an absolute calibration of the autocollimator scale. Calibration runs were taken
over a period spanning 85 days. {\bf Bottom plot:} power spectral density of the torque signal at $s\!=\!72\,\mu$m. Smooth lines show the thermal (and autocollimator) noise assuming $Q=1000$. The $54\omega$ peak is the 3rd harmonic of the $n\!=\!18$ signal. For this run, the $18\omega$ and $120\omega$ signals were set to 1/2 and 10/3 times the free-oscillation frequency $\omega_0$. The $36\omega$ peak occurs at $\omega_0$ and is a residue of the free oscillation; it is not a 2nd harmonic torque of the $n\!=\!18$ pattern as all even harmonics vanish in our geometry.}
\label{fig2}
\end{figure}
The main challenges were fabricating and then positioning with $\mu$m accuracy 5.5 cm-diameter objects (one of them suspended from an 83 cm-long torsion fiber),  minimizing stray electrostatic and magnetostatic effects, dealing with seismic vibrations, and eliminating dust particles that can prevent taking data at small $s$.  The detector was gold-coated and surrounded by a rigid, almost hermetic, gold-coated electrostatic shield; its key element was a $10\mu$m-thick, tightly-stretched gold-coated beryllium-copper foil located between the detector and attractor.  Seismic vibrations coupled to patch fields substantially increased the noise when the pendulum-foil separation was less than $30\, \mu$m. Multi-layer $\mu$metal shields isolated the detector from external fields as well as from the turntable motor. Instrumental temperature variations during a run were controlled at the $\pm 5~$mK level.

In our first generation work\cite{co:13}, the test bodies were cut from $50~\mu$m thick tungsten foils by electric discharge machining and were kept flat by attaching them to Pyrex glass annuli using Dow Integral E100 adhesive film. They were then coated with gold and mounted on the pendulum frame and attractor turntable. The much smaller test body scale compared to our earlier work\cite{ho:04,ka:07} required new instrumentation (SmartScope\cite{OGP}) for characterizing and aligning the test bodies, more precise turntable control, and new electrostatic shields that provided better access for removing dust particles. Otherwise the instrument and general principles of the analysis were the same as in Ref.~\cite{ka:07}.  
This work resolved at $\approx 50 \sigma$ the $18\omega$ gravitational signal between two objects with masses of only 200 mg. To our knowledge, this was easily the smallest mass-scale for which the gravitational interaction had been resolved\cite{sc:16}.  We did not publish that result, which probed separations between $57~\mu$m and 2.00 mm, because there were hints of a subtle systematic effect that we were unable to identify.
%
%
\begin{figure}[!]
\includegraphics[width=.40\textwidth]{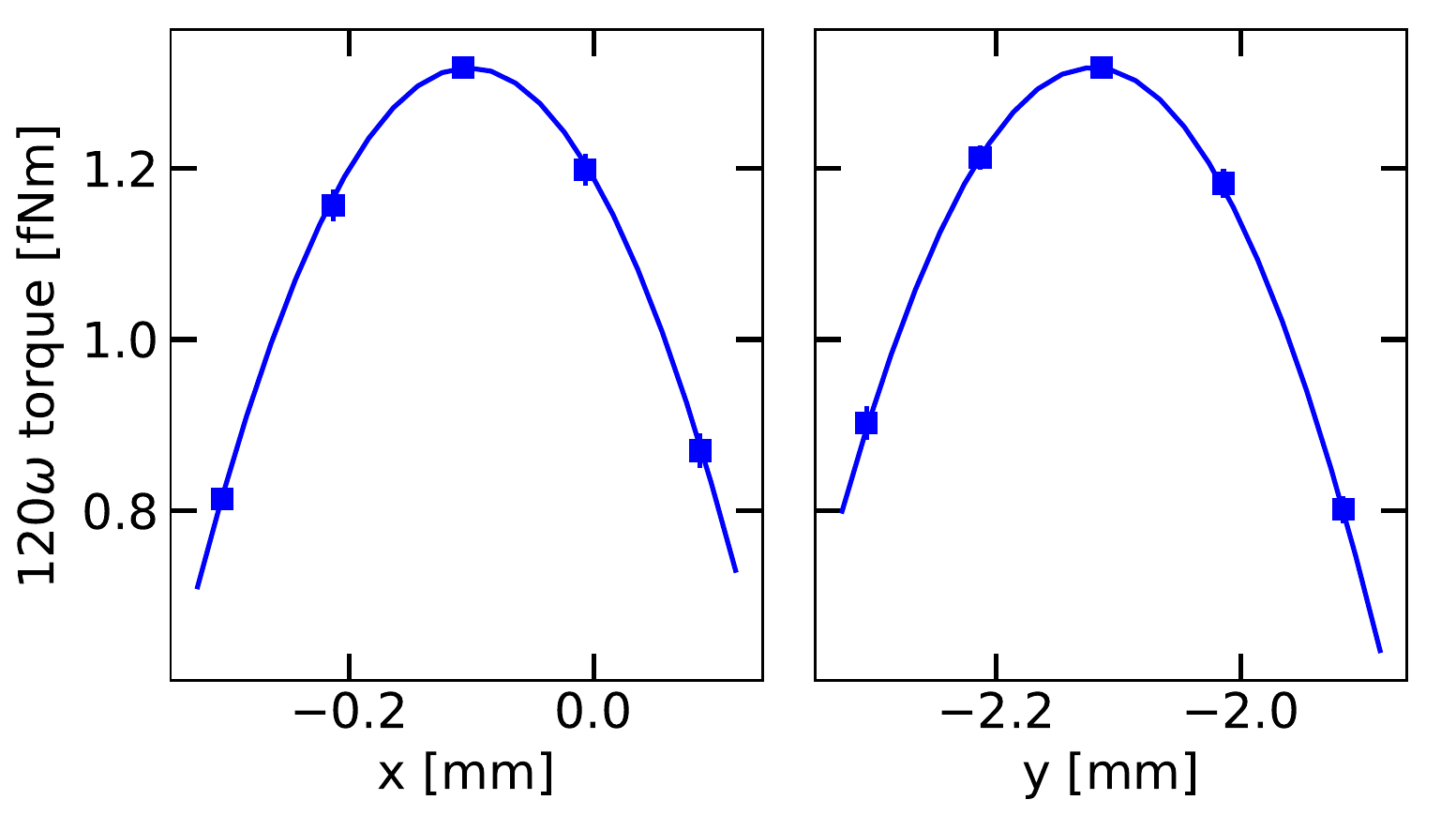}     
\includegraphics[width=.40\textwidth]{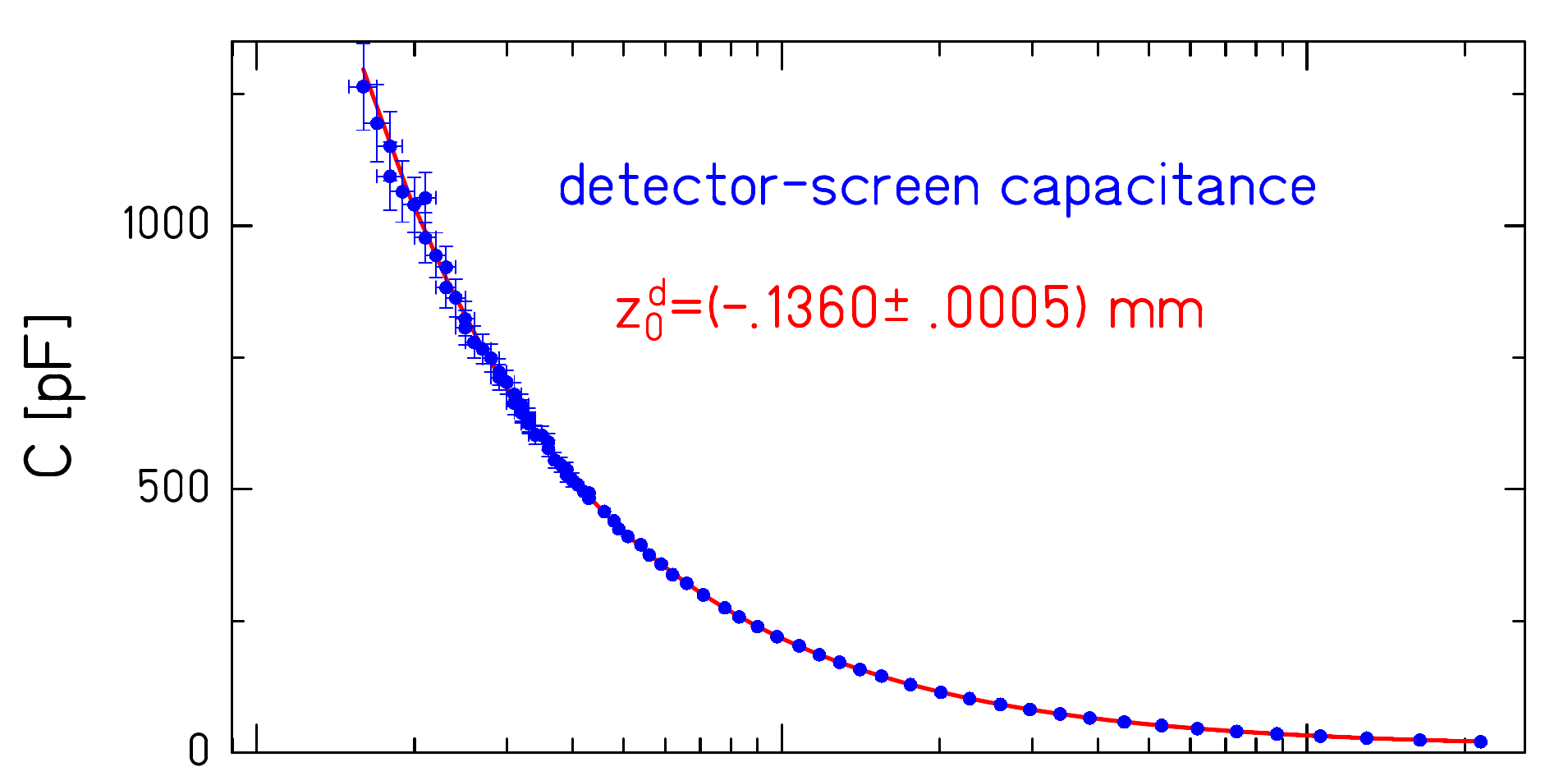}        
\includegraphics[width=.40\textwidth]{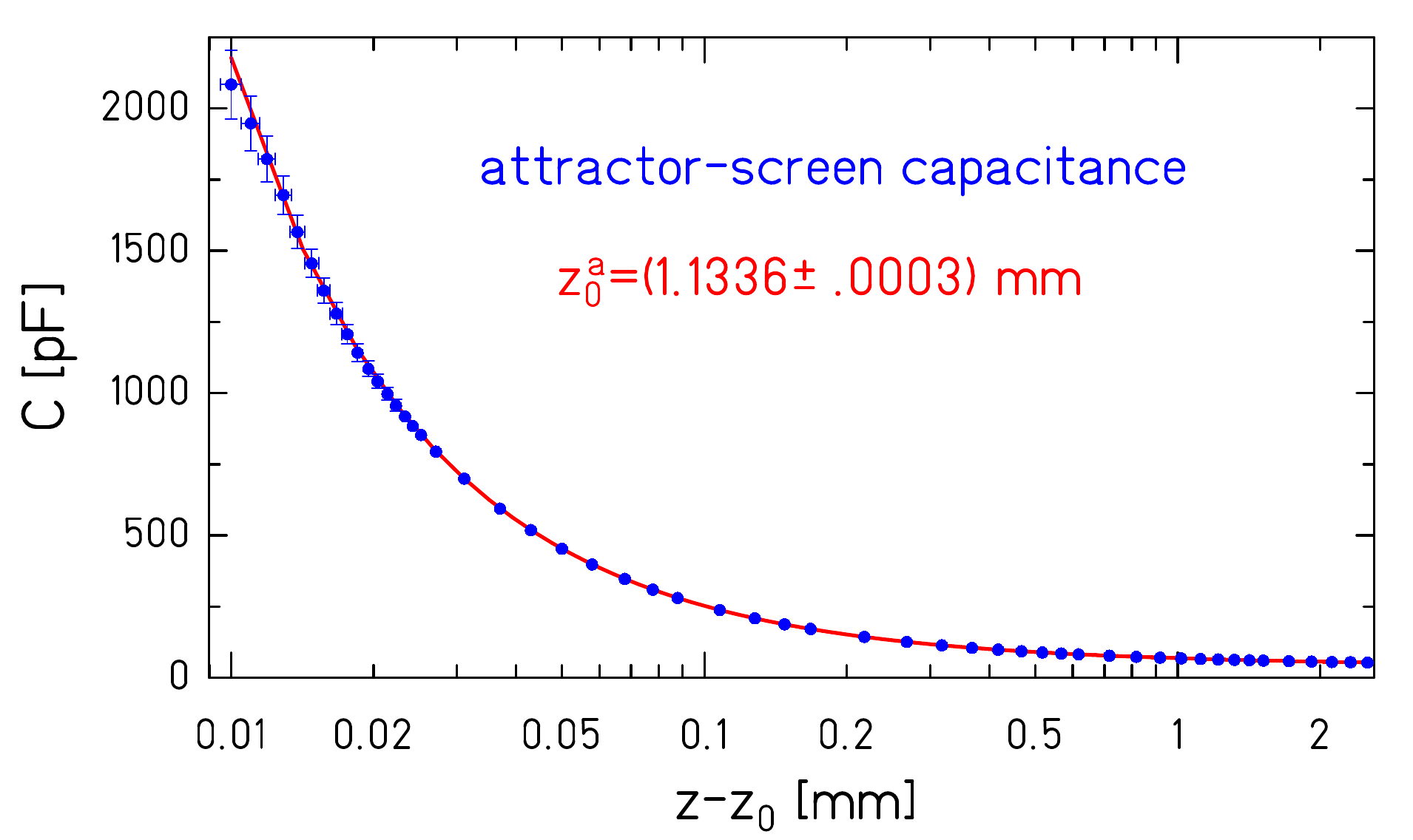}        
\caption{Generation 2 data. {\bf Top plot:}  horizontal centering of the detector on the attractor using the gravitational $120\omega$ torque. The center occurs at $x_0$=$(-102\pm 2)\,\mu$m, $y_0$=$(-2121\pm 2)\,\mu$m.
{\bf Middle and bottom plots:} the vertical separation $s$ was determined by combining two measurements of electrical capacitance plus the thicknesses of the foil and glue films on the test body faces. When error bars are not in shown this and later figures, the uncertainties are smaller than the symbols.}
\label{fig3}
\end{figure}

Our second generation work\cite{le:19}, whose results we present here, used platinum test bodies with detector and attractor thicknesses of 54 and 99 $\mu$m, respectively. These were epoxied to BK7 glass annuli using a technique, similar to that described in Ref.~\cite{we:08}, that filled the hole pattern with glue so that the test-body faces were flat to within $\pm 2.3~\mu$m and $\pm 1.5~\mu$m. In addition, we modified the vacuum vessel and pumping system to accommodate an {\em in situ} system for remotely positioning the $10~\mu$m-thick isolation foil with sub-$\mu$m accuracy. We took data at separations between $52\,\mu$m and $3.0\,$mm.

\begin{figure}[!t]
\includegraphics[width=.37\textwidth]{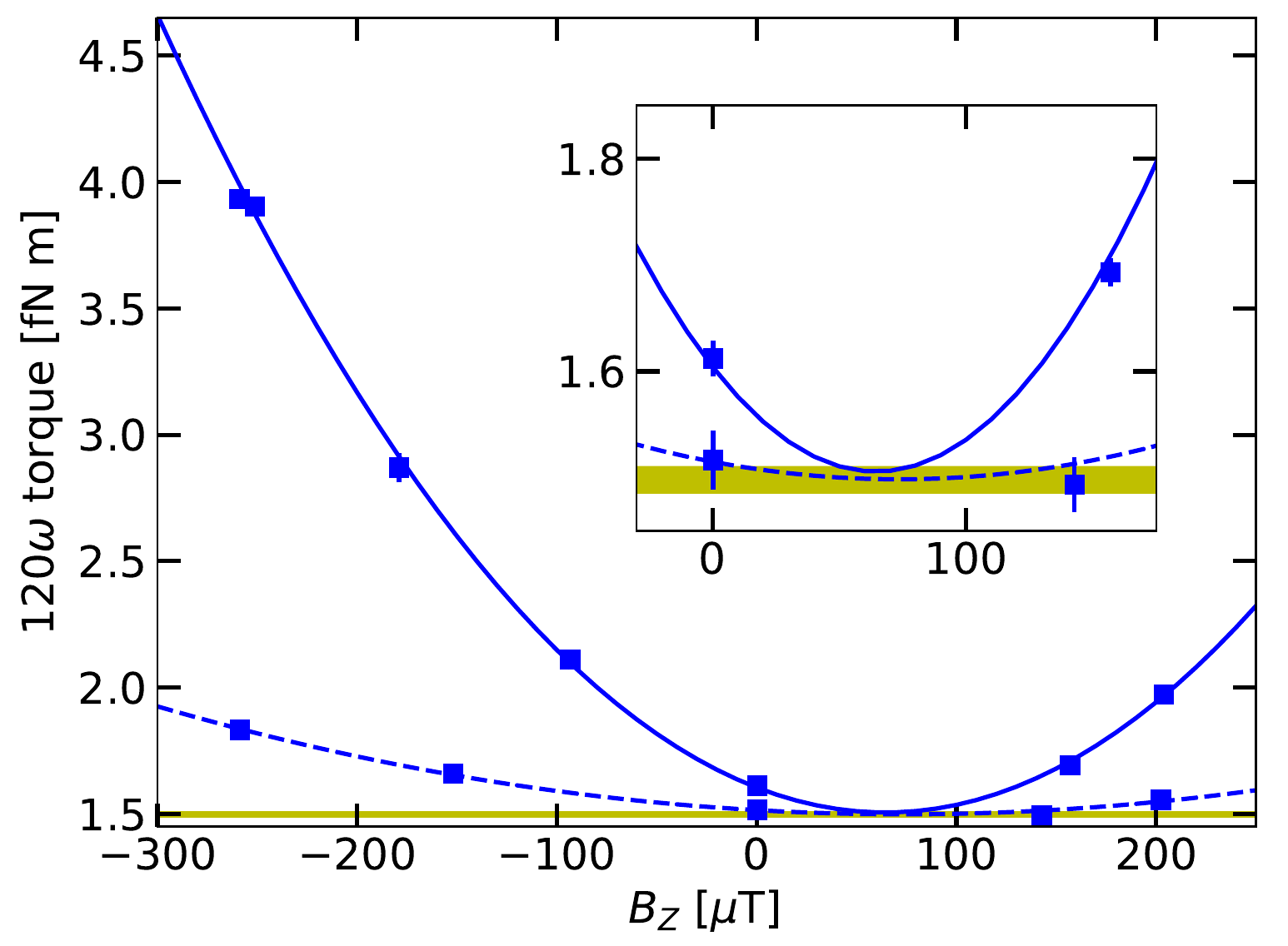}      
\includegraphics[width=.38\textwidth]{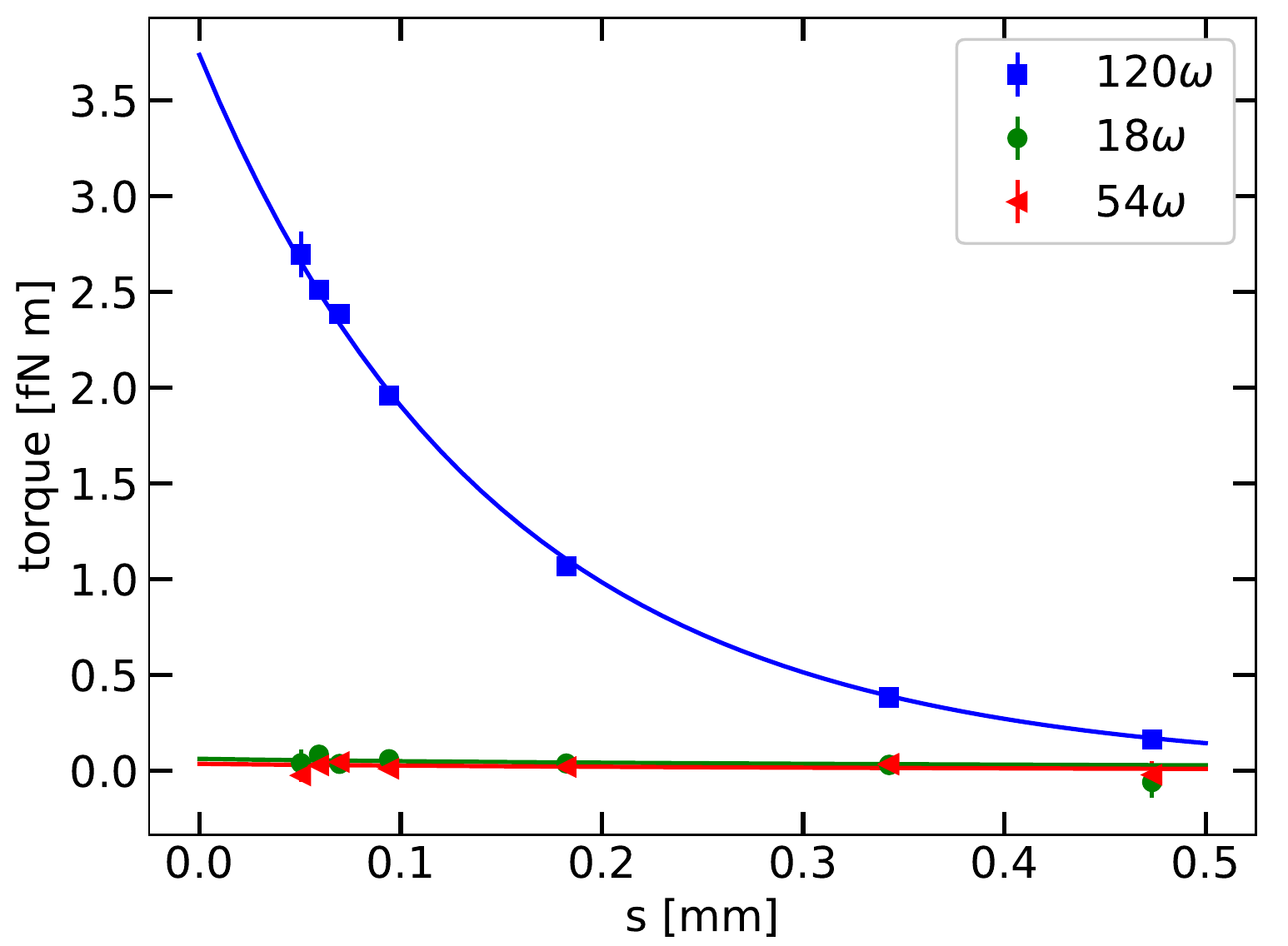}      
\caption{(color online)  {\bf Top plot:} effect  of an applied vertical magnetic field $B_z$ at $s\!=\!72\,\mu$m.  Points on the solid and dashed lines were taken with the outermost magnetic shield removed and in place, respectively. The lines are parabolic fits. The field at the detector vanishes at $B_z=65~\mu$T. The horizonal green band shows the the measured $s\!=\!72\,\mu$m torque in our science data. The ($.0165\!\pm\! .0054)\,$fN\,m difference between the $B_z\!=\!0$ and $B_z\!=\!65\,\mu$T torques is the magnetic contribution to the torque.   
{\bf Bottom plot:} Systematic effect of a $B_z$ field as a function of $s$. Points show the difference between torques at a strong applied field ($B_z=-250~\mu$T: outermost shield removed) and a nulled field ($+65~\mu$T: all shields in place). The smooth curves show our Fourier-Bessel calculations of the spin-spin interaction between the induced magnetizations in the detector and attractor test bodies; a single normalization reproduces the $18\omega$, $54\omega$ and $120\omega$ effects.}
\label{fig4}
\end{figure}

\begin{figure}[b]
\includegraphics[width=.46\textwidth]{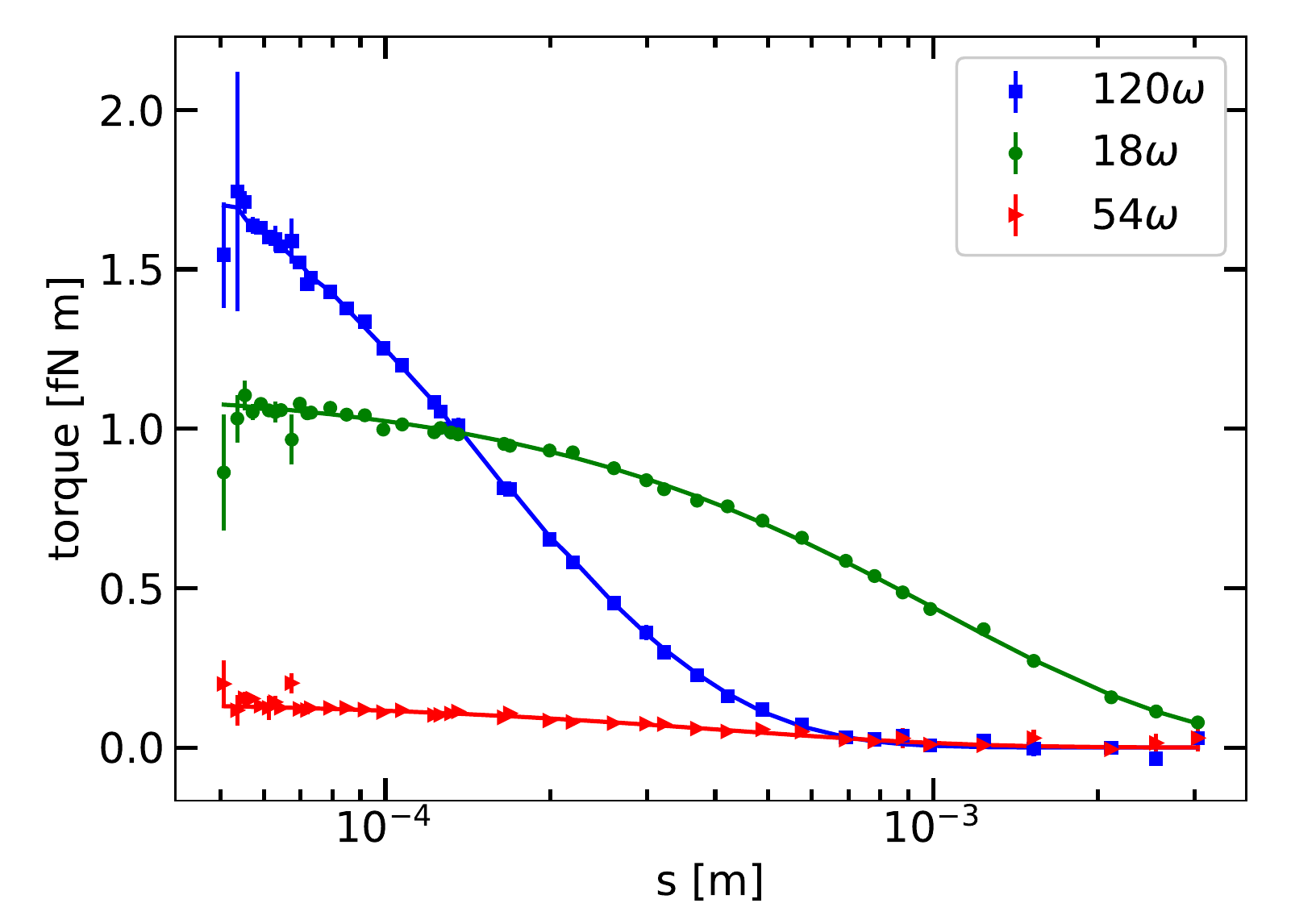}	      
\includegraphics[width=.48\textwidth]{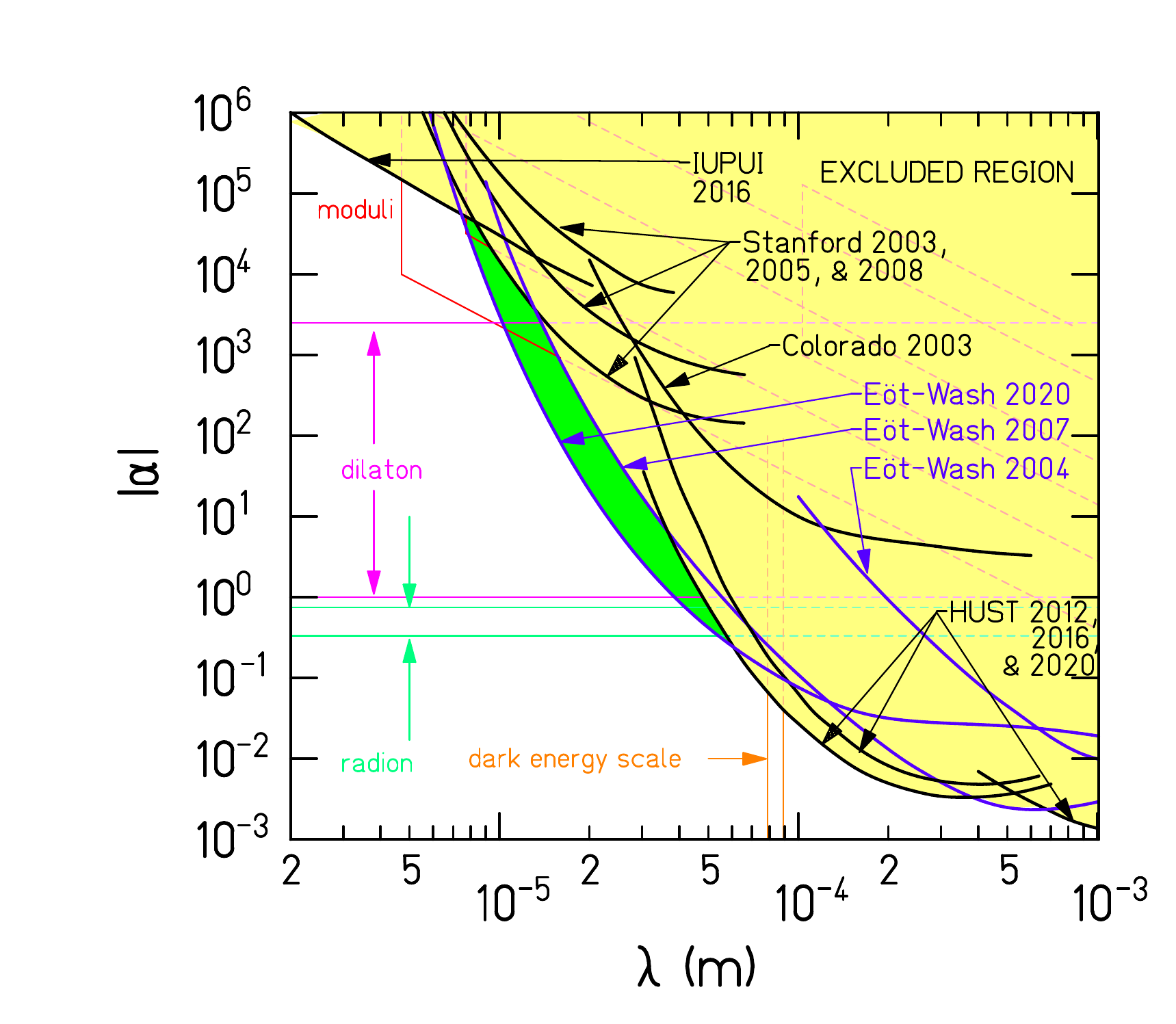}		
\caption{(color online) {\bf Top plot:} 18$\omega$, 54$\omega$ and 120$\omega$ torques corrected for the magnetic systematic. Unless shown otherwise, uncertainties are smaller than the plot symbols. Data points with essentially the same $s$ are combined for display  purposes only. The Newtonian fit is shown and has $P\!=\!0.654$. Adding a Yukawa term did not improve the fit appreciably. {\bf Bottom plot:}  corresponding 95\% confidence upper limits on $|\alpha|$ from this and previous works\cite{ka:07,ho:04,ho:85,lo:03,ch:03,sm:05,ge:08,huaz:12,huaz:16,ta:20}.}
\label{fig5}
\end{figure}

We made the usual checks for systematic effects\cite{ho:04,ka:07} by varying temperatures and their gradients, impressing electrical and magnetic fields on the detector and attractor, {\em etc}.
The only significant systematic effect arose from the magnetic susceptibility of the platinum test bodies. An ambient magnetic field will slightly magnetize the metal test bodies inducing a dipole-dipole interaction between the attractor and detector that is not attenuated by our electrostatic shield. The induced magnetic interaction from a vertical field $B_z$ is attractive, while for horizontal fields $B_x$ and $B_y$ the induced interaction is repulsive.
 We studied this systematic effect using Helmholtz coils and observed just such behavior. The largest effect arises from $B_z$ fields and is shown in the top panel of  Fig.~\ref{fig4}.
%
The effects of horizontal fields were at least 20 times smaller. With the detector at  $s\!=\!72\,\mu$m we observed the  $120\omega$ torque as a function of $B_z$. This torque was parabolic with a minimum at $B_z=+65~\mu$T and a strength consistent with the value measured in our science data. Furthermore, the  $+65~\mu$T value was consistent with that needed to cancel the measured ambient field. This demonstrated that, as expected, there was no magnetic effect linear in $B_z$, and it allowed us to determine the magnetic contribution to the  $120\omega$ torque. 
Then, as shown in the bottom panel of Fig.~\ref{fig4}, we measured the $s$-dependence of the magnetic $18\omega$, $120\omega$ and  $54\omega$ (the 3rd harmonic of the $18\omega$ signal) torques.  The Fourier-Bessel framework provides semi-analytic solutions for the magnetic dipole-dipole\cite{te:15} torques in azimuthal\cite{te:15} or axial\cite{le:19} geometries.
Fourier-Bessel spin-spin predictions with a single adjustable scale parameter agree with our observations. The
$120\omega$ signal had the only significant (roughly 1\%) magnetic contribution which we subtracted using the data in Fig.~\ref{fig4}. The resulting ``gravitational'' torques are shown in Fig.~\ref{fig5}.

We fit our $j=95$ measurements of $m=3$ torques shown in the top panel of Fig.~\ref{fig5},
$N_m(\zeta_j)\pm \delta N_m(\zeta_j)$, with predicted torques,  $\tilde{N}_m(\vec{\zeta}_j,\vec{\eta},\lambda)$, that were functions of 17 experimental parameters, $\vec{\eta}$. These were constrained to $\eta_i^{exp}\pm \delta\eta_i^{exp}$ using micrometers, measuring microscopes, electronic scales, {\em etc}. The errors in 12 parameters (4 masses and 4 thicknesses of the material removed to create the 18-fold and 120-fold hole patterns in the detector and attractor test bodies, the density of the glue that filled the holes, 
SmartScope measurements of attractor runout and tilt, plus capacitance measurements\cite{co:13} of pendulum tilt) were so well constrained that the corresponding uncertainties in the predicted torques were negligible. The uncertainties in 5 parameters, $x_0, y_0$ and $s_0$  (the sum of the thicknesses of the isolation foil and the glue films on the faces of the detector and attractor test bodies), 
a surface-roughness correction, and the autocollimator angle scale $\gamma$,  had noticeable effects of the predicted torques. We allowed those parameters to float in fitting our torques but added terms to $\chi^2$ to constrain them by SmartScope measurements and the data shown in Figs. \ref{fig2} and \ref{fig3}.
We accounted for the uncertainties in $\zeta$ by minimizing  
\begin{displaymath}
\chi^2(\lambda) \!=\!\!\sum_{j=1}^{95}\!\sum_{m}\!\frac{[N_m(\vec{\zeta_{j}})\!-\!\tilde{N}_m(\vec{\zeta_{j}},\vec{\eta}, \lambda)]^2}{(\delta N_m)^2\!+\! \left(\delta s_j\frac{\partial\tilde{N}_m}{\partial s_j}\right)^2} \!\!+\! \sum_{i=1}^{5}\!\bigg[\frac{\eta^{exp}_{i}\!-\!\eta_{i}}{\delta \eta^{exp}_{i}}\bigg]^2
\end{displaymath}
where $\delta s_j$ is the error in $s$ arising from uncertainties in the measured capacitances.
%

We first tested the Newtonian model ($\lambda\!=\!\infty$) and, as shown in the top panel of Fig.~\ref{fig5},  obtained an excellent fit: $\chi^2\!=\!275.0$ with $\nu=285$.  
We then tested for a single additional Yukawa term by finding the constraints on $\alpha$ for 66 assumed values of  $\lambda$ between $5\,\mu$m and $9\,$mm; $\vec{\eta}$ was allowed to vary independently at each $\lambda$. None of these Yukawas improved $\chi^2$ at the 2$\sigma$ level ($\Delta \chi^2 = 6.2$); the best fit $\Delta\chi^2\!=\!3.3$ occurred for $\lambda\!=\!7.1\mu$m. The bottom panel in Fig.~\ref{fig5}
displays our $2\sigma$  constraints on $|\alpha|$ (constraints on $+\alpha$ and $-\alpha$ are given in Supplemental Material\cite{sup}). We find that any gravitational-strength Yukawa interaction must have $\lambda<38.6\,\mu$m. 
This implies that the dilaton\cite{ka:00} or heavy graviton\cite{ao:16} mass, and the radion unification\cite{ad:03,ad:07} mass must be greater than 5.1 meV and 7.1 TeV, respectively, and that the largest extra dimension\cite{ad:03} must have a toroidal radius less than 30 $\mu$m. These are the tightest existing lab constraints on ``string inspired'' new gravitational phenomena.   

Environmental vibrations prevented us from probing separations smaller than $52 \mu$m and increased the noise in the smaller separation data. We are now implementing an active system to reduce vertical vibrations.

W.\,J. Kim helped commission the foil-positioning mechanism; C.\,A. Hagedorn provided thoughtful advice throughout.
This work was supported in part by National Science Foundation Grants PHY-1305726, PHY-1607391 and PHY-1912514. 

\end{document}